\documentclass[%
 aip,
 amsmath,
 amssymb,
 reprint
]{revtex4-2}

\usepackage{graphicx}% Include figure files
\usepackage{dcolumn}% Align table columns on decimal point
\usepackage{bm}% bold math
\usepackage[utf8]{inputenc}
\usepackage[T1]{fontenc}
\usepackage{xcolor,color,soul}
\usepackage{mathptmx}
\usepackage{etoolbox}
\usepackage{upgreek} 
\usepackage{multirow}
\usepackage{algorithm2e}
\makeatletter
\def\frontmatter@thefootnote{%
 \altaffilletter@sw{\@fnsymbol}{\@fnsymbol}{\csname c@\@mpfn\endcsname}%
}%
\makeatother

\usepackage{colortbl}
\usepackage{hyperref}

\makeatletter
  \def\href@noop#1#2{#2}
  \let\href\href@noop
\makeatother

\begin{document}

% \preprint{AIP/123-QED}

\title[]{Transformer-Based Inverse Microrheology for Experimental Mechanics at Ultra-High Strain Rates} 
%  Using Laser-Induced Cavitation and Machine Learning-Based Property Inference

% Force line breaks with \\

\author{Lehu Bu } 
\affiliation{Materials Science Graduate Program, Texas Materials Institute, The University of Texas at Austin, Austin, TX, USA, 78712} 

\author{Zhaohan Yu}
%\email{yuzhaoha@msu.edu}
\affiliation{Department of Mechanical Engineering, Michigan State University, East Lansing, MI, USA, 48824} 

\author{Danila Frolkin} 
\affiliation{Department of Aerospace Engineering and Engineering Mechanics, The University of Texas at Austin, Austin, TX, USA, 78712} 

\author{Junyoung Kim} 
\affiliation{Department of Aerospace Engineering and Engineering Mechanics, The University of Texas at Austin, Austin, TX, USA, 78712} 

\author{Qihang Shi} 
\affiliation{Department of Aerospace Engineering and Engineering Mechanics, The University of Texas at Austin, Austin, TX, USA, 78712}

\author{Jan N. Fuhg}
\affiliation{Department of Aerospace Engineering and Engineering Mechanics, The University of Texas at Austin, Austin, TX, USA, 78712} 
\affiliation{The Oden Institute of Computational Science and Engineering, University of Texas at Austin, Austin, TX, USA,  78712}

\author{Shaoting Lin }
%\email{linshaot@msu.edu}
\affiliation{Department of Mechanical Engineering, Michigan State University, East Lansing, MI, USA, 48824}
\affiliation{Department of Mechanical Engineering, University of Wisconsin-Madison, Madison, WI, USA,53706}

\author{Jin Yang}
\email{jin.yang@austin.utexas.edu; Corresponding author}
\affiliation{Materials Science Graduate Program, Texas Materials Institute, The University of Texas at Austin, Austin, TX, USA, 78712} 
\affiliation{Department of Aerospace Engineering and Engineering Mechanics, The University of Texas at Austin, Austin, TX, USA, 78712}

 %\altaffiliation[]{Department of Aerospace Engineering \& Engineering Mechanics, The University of Texas at Austin} %Lines break automatically or can be forced with \\

%\author[add1]{Jin~Yang}
%\author[add1]{Alexander~McGhee}
%\author[add1]{Griffin~Radtke}
%\author[add1]{Christian~Franck\corref{cor}}
%\cortext[cor]{Corresponding author. E-mail address: cfranck@wisc.edu (C. Franck)} 
%\address[add1]{Department of Mechanical Engineering, University of Wisconsin-Madison, Madison, WI, USA}

\date{\today} % It is always \today, today,

\begin{abstract}
Traditional rheological tools are often limited in characterizing soft materials under ultra-high strain-rate loading conditions ($> 10^3$ s$^{-1}$) due to constraints in spatiotemporal resolutions, loading rate, and invasiveness. Recently, inertial microcavitation rheometry (IMR), which utilizes laser-induced inertial cavitation (LIC) to dynamically deform surrounding materials, has emerged as a powerful experimental mechanics technique for probing nonlinear viscoelastic properties under extreme loading conditions. However, conventional IMR relies on computationally expensive iterative inverse fitting procedures, limiting its scalability and real-time applicability.
Here, we introduce a new AI-enhanced experimental mechanics framework, called {Bubble Dynamics Transformer} (BDT), that integrates physics-based cavitation simulations with Transformer neural network architectures to achieve rapid inverse characterization of soft material viscoelasticity from experimentally measured bubble dynamics. The proposed framework directly predicts viscoelastic material parameters from time-resolved bubble radius evolution curves without iterative optimization. The BDT is trained using synthetic datasets generated from physics-based Keller–Miksis cavitation simulations and validated using experimental laser-induced cavitation data obtained from hydrogels and viscous polymer solutions. The proposed AI-driven framework demonstrates excellent agreement with our previous IMR while substantially accelerating constitutive parameter inference. Experimental demonstrations further reveal the capability of the framework to characterize rate-dependent material behavior across a wide range of soft materials, from viscous liquids to various viscoelastic hydrogels,   at ultra-high strain rates.

{  } 

\end{abstract}
 
\maketitle

% =================================
\section{Introduction}
\label{sec:intro}
% =================================
The rheological characterization of soft materials at ultra-high strain rates ($> 10^{3}$ s$^{-1}$) is important for the understanding of their dynamic mechanical behavior across a wide spectrum of engineering and biomedical applications. For instance, precise characterization of tissue properties during rapid loading is important for ensuring the safety and efficacy of laser and ultrasound-based surgical interventions while minimizing unwanted collateral damage \cite{brennen2015cavitation}. 
However, conventional techniques such as oscillatory shear plate rheometry \cite{hyun2011review}, dynamic mechanical analysis (DMA) \cite{chartoff2009dynamic}, and micro- or nano-indentation \cite{dufrene2008towards,oliver2010nanoindentation} often fall short when confronted with these extreme loading conditions, particularly those encountered in ultra-high strain-rate regimes where strain rates exceeding 10$^3$ s$^{-1}$.

Recently, inertial microcavitation rheometry (IMR) \cite{estrada2018high,yang2020extracting,yang2022mechanical,yang2024estimating,bremer2024ballistic,zhu2025parsimonious} has emerged as a promising and minimally invasive experimental technique to address this critical need. By focusing a femtosecond- or nanosecond-duration laser pulse or a high-intensity ultrasound pulse within a transparent soft material, such as a liquid, hydrogel, elastomer, or biological tissue, IMR initiates a localized microscale cavitation bubble above a critical energy threshold. The subsequent rapid radial expansion and collapse of this bubble dynamically load the surrounding material at ultra-high, ballistic strain rates, enabling localized and extreme-rate rheological probing. By capturing the temporal evolution of the bubble dynamics, for example, the bubble radius $R(t)$, using ultra-high-speed imaging and comparing these measurements with theoretical cavitation simulations, the constitutive properties of the surrounding material can be inferred.

IMR has been successfully applied to characterize a variety of viscous liquids $\cite{zhu2025parsimonious}$ and hydrogels, including polyacrylamide \cite{estrada2018high,yang2020extracting,zhu2025parsimonious,sanchez2026hierarchical}, agarose \cite{yang2022mechanical}, gelatin \cite{mcghee2023high,bremer2024ballistic,sanchez2026hierarchical}, fibrin hydrogel \cite{luo2020laser}, and collagen \cite{sreedasyam2026scanning}. Despite these successes, conventional IMR-based constitutive parameter identification remains computationally expensive because the forward cavitation model must be repeatedly solved during iterative inverse optimization procedures.

More broadly, recent advances in artificial intelligence (AI) and scientific machine learning are rapidly transforming experimental mechanics by enabling intelligent interpretation of experimental datasets, accelerating inverse characterization workflows, and integrating experimental measurements with data-driven constitutive modeling. In particular, AI-enhanced inverse experimental mechanics frameworks offer the potential to bypass computationally expensive iterative optimization procedures traditionally required for constitutive parameter identification. However, the integration of AI with extreme-rate experimental mechanics and cavitation-driven microrheology remains largely unexplored.

%%%%%%%%%%%%%%%%%%%%%%%%%%%%%%%%%%%%%%%%%%%%
\begin{figure*}[t!]
    \centering
    \includegraphics[width=0.9 \textwidth]{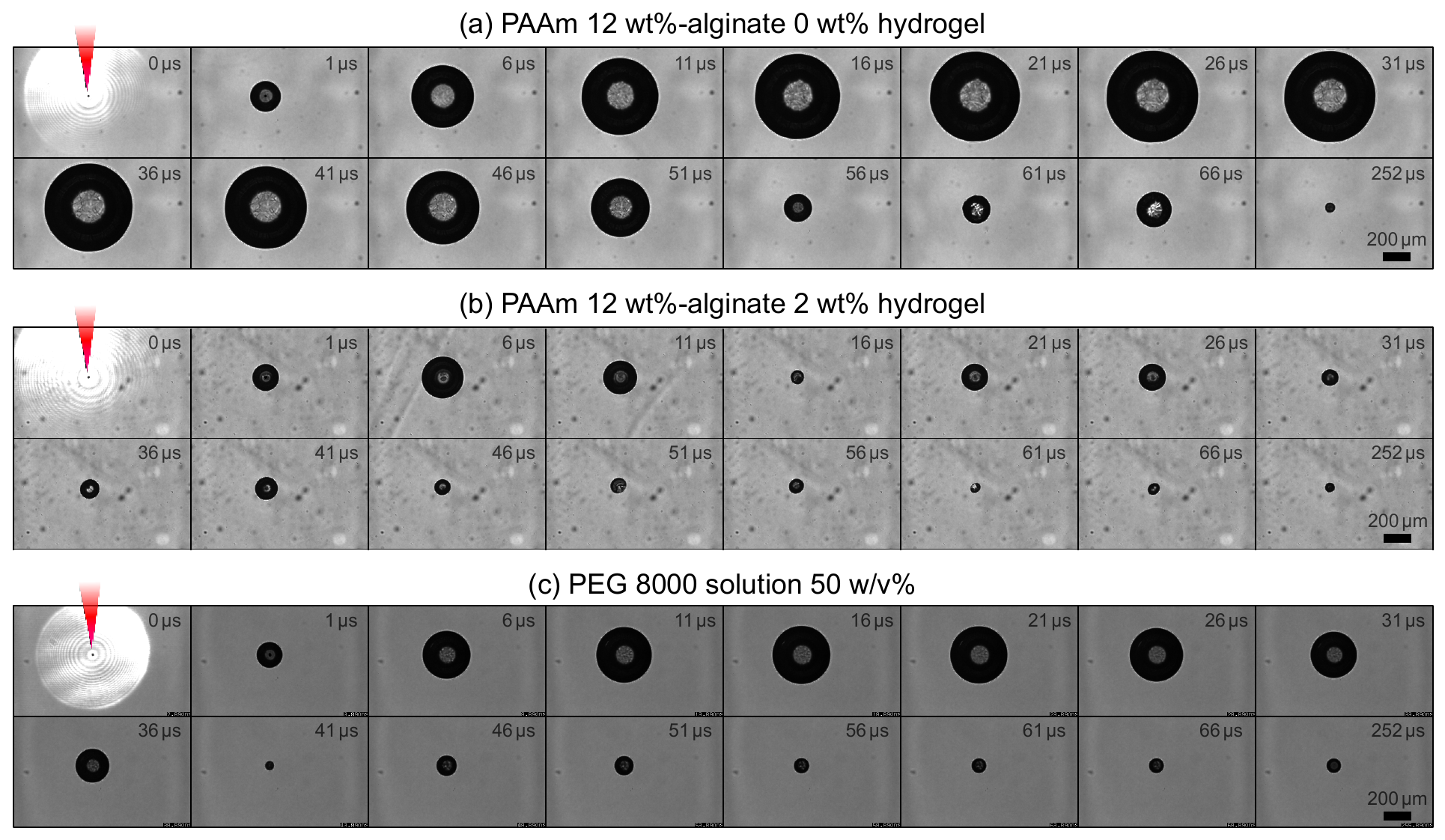}
    \caption{Representative high-speed image sequences (selected frames) of laser-induced cavitation dynamics in soft materials:
(a) PAAm 12 wt\%-alginate 0 wt\% hydrogel, (b) PAAm 12 wt\%-alginate 2 wt\% hydrogel, and (c) PEG 8000 solution at 50 w/v\%. Original ultra-high-speed images were recorded at 1 million frames per second.}
    \label{fig: LIC img seq}
\end{figure*}
%%%%%%%%%%%%%%%%%%%%%%%%%%%%%%%%%%%%%%%%%%%%

In this work, we introduce a new AI-enhanced experimental mechanics framework that integrates physics-based cavitation modeling with Transformer neural network architectures \cite{vaswani2017attention}. Specifically, we develop a physics-informed deep learning framework, termed the Bubble Dynamics Transformer (BDT), that directly infers constitutive material properties from experimentally measured cavitation bubble dynamics. By leveraging synthetic training datasets generated from physics-based cavitation simulations, the proposed framework enables rapid inverse characterization of soft material viscoelasticity without requiring computationally expensive iterative fitting procedures.

% =================================
\section{Methodology }
\label{sec:method}
% =================================

\subsection{Experimental Setup} 
\label{sec:exp}

%%%%%%%%%%%%%%%%%%%%%%%%%%%%%%%%%%%%%%%%%%%%
\begin{figure*}[t!]
    \centering
    \includegraphics[width=0.9 \textwidth]{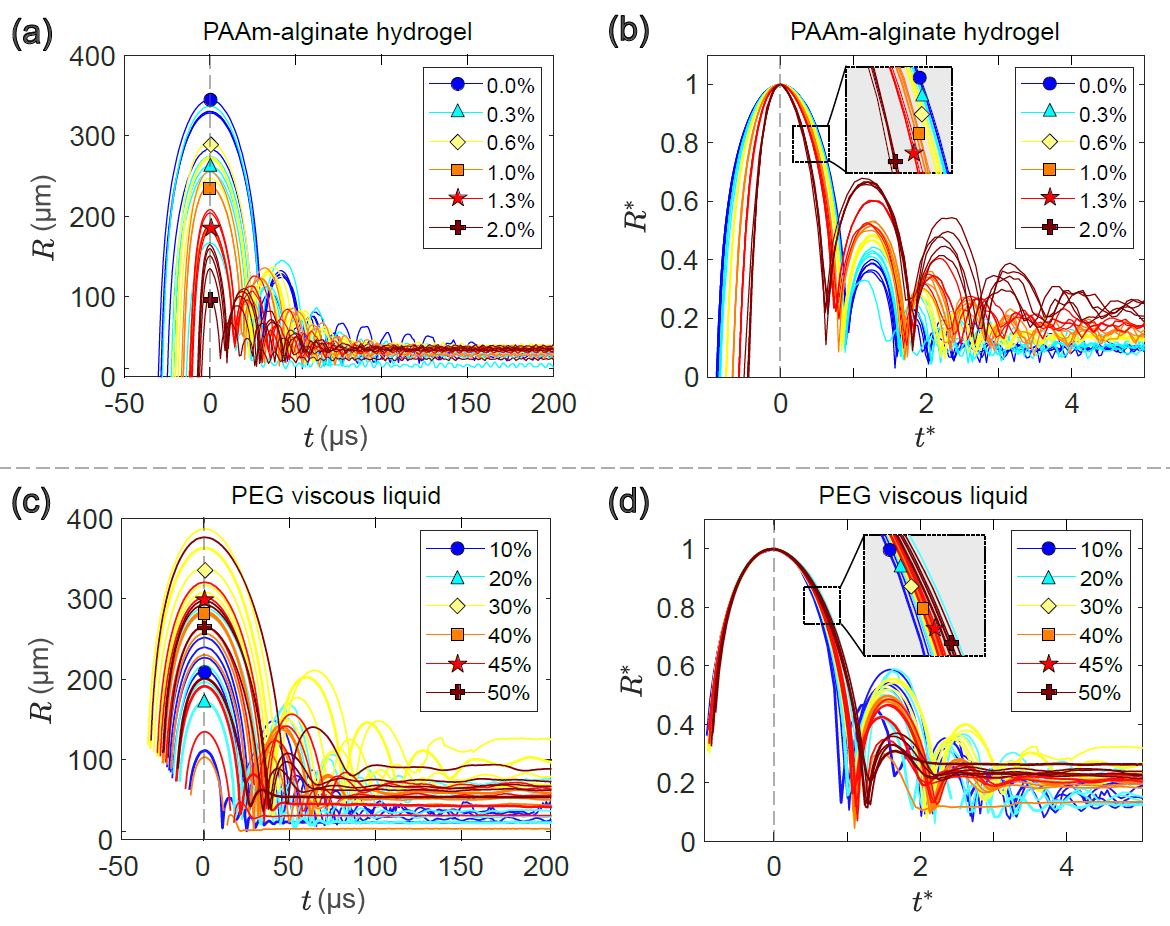}
    \caption{(a)~Dimensional radius $R$  vs. time $t$ curves for laser-induced inertial cavitation (LIC) experiments in ``12 \% PAAm + X \% alginate'' double network hydrogels (X$\%$ denotes the concentration of alginate indicated in the figure legend). 
    (b)~Nondimensionalized bubble radius $R^*$ vs. time $t*$ curves for data shown in (a).
    (c)~Dimensional radius $R$ vs.\ time $t$ curves for laser-induced inertial cavitation (LIC) experiments in PEG 8000/water mixtures with varying PEG 8000 concentrations (10--50~w/v\%, as indicated in the figure legend).
    (d)~Nondimensionalized bubble radius $R^*$ vs. time $t*$ curves for the data shown in (c).}
    \label{fig: Rt curves}
\end{figure*}
%%%%%%%%%%%%%%%%%%%%%%%%%%%%%%%%%%%%%%%%%%%%

In IMR, inertial microcavitation events can be induced through focused laser pulses \cite{liang2022comprehensive} or high-intensity ultrasound pulses \cite{mancia2021acoustic}. In this work, we focus on LIC because pulsed laser excitation can generate a rapid bubble expansion and collapse with high location precision. The deformation field generated by the bubble is inherently non-uniform in both space and time. Material elements near the bubble wall experience very high strain rates during collapse, whereas larger strain amplitudes occur near the maximum bubble expansion. Both strain and strain rate decay as the bubble approaches its equilibrium radius, and material points farther away from the bubble wall experience substantially smaller deformation. In this work, the reported strain rate is defined as the average strain rate between the maximum bubble radius and the first violent collapse. Although cyclic ultrasound excitation can drive bubble oscillations at prescribed frequencies, reaching the high-strain and high-strain-rate regimes characteristic of IMR is more challenging using this approach \cite{saint2020acoustic,murakami2021ultrasound}.

% LIC exp setup
%\noindent \textbf{Laser-induced cavitation (LIC): }
In the LIC experiments, a single 3-5~ns pulse from an adjustable 1–25~mJ Q-switched Nd:YAG Minilite II laser (Continuum, Milpitas, CA), frequency-doubled to 532~nm or operated at 1064~nm, is focused within the sample. The resulting cavitation bubble dynamics were recorded at frame rates of 1–10 million frames per second using an ultra-high-speed camera (HPV-X2, Shimadzu, Japan). Additional details of the LIC experimental configuration can be found in our previous work \cite{estrada2018high,yang2020extracting}. 
Figure~\ref{fig: LIC img seq} shows representative high-speed image sequences of LIC dynamics in the viscoelastic solids and viscous liquids investigated in this study.

IMR has previously been applied to characterize various soft material systems  \cite{estrada2018high,yang2019strain,yang2020extracting,yang2022mechanical,mcghee2023high,bremer2024ballistic,estrada2021neural,luo2020laser,sreedasyam2026scanning,sanchez2026hierarchical}.  
In this work, PAAm-alginate double-network hydrogels \cite{lin2022extreme} were selected as model viscoelastic solids. We also test the developed BDT deep learning framework using other previously studied hydrogels. The experimental protocols for fabricating PAAm-alginate hydrogels are summarized in Lin et al. \cite{lin2022extreme}. Briefly, PAAm-alginate hydrogels consist of two interpenetrating polymer networks: a covalently crosslinked long-chain PAAm network and an ionically crosslinked short-chain alginate network. The PAAm network provides stretchable elasticity, whereas the alginate network can dissociate under large deformation, thereby contributing to energy dissipation, strain stiffening, and enhanced fracture resistance. 
Figure~\ref{fig: Rt curves}(a) shows experimentally measured bubble radius--time curves, $R(t)$, obtained from ``12\% PAAm + X\% alginate'' double-network hydrogels, where X denotes the alginate concentration. The corresponding normalized curves are shown in Fig.~\ref{fig: Rt curves}(b), where $R^* = {R}/{R_{\max}}$,  $t^* = {t}/{t_c}$.
Here,  $R_{\max}$ is the maximum bubble radius. $t_c$ is defined as $R_{\max}/U_c$ where  $U_c = \sqrt{p_{0}/\rho}$; $\rho$ is surrounding material mass density ($\rho \sim 1,000 \ \text{kg}/\text{m}^3$); $p_0$ is the far-field ambient pressure ($p_0 \sim$ 101,325 Pa). \\

To further evaluate the proposed IMR framework in viscous liquids, polyethylene glycol (PEG) 8000 solutions were prepared using PEG 8000 with a molecular weight of 8000~g/mol (Rigaku Reagents; 50\% w/v stock solution) \cite{zhu2025parsimonious}. The stock solution was diluted with deionized water to obtain PEG 8000/water mixtures with concentrations ranging from 10 to 50~w/v\%. Each sample was mixed in a 50~mL centrifuge tube and homogenized using a laboratory shaker for approximately 1~min to ensure uniformity. The measured LIC bubble radius--time curves for PEG 8000/water mixtures are shown in Fig.~\ref{fig: Rt curves}(c), and the corresponding normalized curves are shown in Fig.~\ref{fig: Rt curves}(d).

Oscillatory shear rheometry was performed to provide low-frequency linear viscoelastic benchmarks for the tested materials. The rheometer measurements and corresponding analysis are summarized in Appendix~\ref{app: rheometer results}.

% =========================================
\subsection{Theoretical Modeling of Microrheology at High Strain Rates}
\label{sec: model}
% =========================================
%
The nucleated bubble dynamics by laser pulses were modeled using a modified Keller--Miksis equation \cite{estrada2018high,yang2019strain,yang2020extracting,tzoumaka2023modeling}. The surrounding material is treated as hyper-viscoelastic and incompressible in the near field, while weak compressibility effects are incorporated through the finite longitudinal wave speed $c$ in the far-field description. The resulting evolution equation for the bubble radius $R$($t$) is given by
% %
% \begin{equation}\label{eq: KM equation}
% \begin{split}
%     \left(1-\frac{\dot{R}}{c}\right)R\ddot{R} & + \frac{3}{2}\left(1-\frac{\dot{R}}{3c}\right)\dot{R}^2  \\ 
%     = &  \frac{1}{\rho}\left(1+\frac{\dot{R}}{c}\right)\left(p_{\rm b} - p_{\infty} - p_{\rm f} - \frac{2\gamma}{R} + S\right)   \\
%     & + \frac{1}{\rho}\dfrac{R}{c}\dot{\overline{\left(p_{\rm b} - p_{\rm f} - \frac{2\gamma}{R} +S\right)}},
%     \end{split}
% \end{equation}
% %

%
\begin{widetext}
\begin{equation}
\label{eq:KM_equation}
\begin{aligned}
\left(1-\frac{\dot{R}}{c}\right)R\ddot{R}
+\frac{3}{2}\left(1-\frac{\dot{R}}{3c}\right)\dot{R}^{2}
&= \frac{1}{\rho}\left(1+\frac{\dot{R}}{c}\right)
\left(p_{\rm b}-p_{\infty} -\frac{2\gamma}{R}+S
\right)+
\frac{1}{\rho}\frac{R}{c}
\frac{\text{d}}{\text{d}t} 
\left(
p_{\rm b} 
-\frac{2\gamma}{R}+S
\right),
\end{aligned}
\end{equation}
\end{widetext}

\noindent where $\rho$ is the mass density of the surrounding soft material, which is assumed to be constant; $\gamma$ is the surface tension of the bubble interface. The overdot denotes differentiation with respect to time. The term $S$ represents the pressure contribution of the stress field in the surrounding hyper-viscoelastic material, which is given by 
\begin{equation}
    S = \int_{R}^{\infty} \frac{2}{r} \left( \sigma_{rr} - \sigma_{\theta \theta} \right) \, \text{d}r, \label{eq: Stress integral}
    \end{equation}
where $\sigma_{rr}$ and $\sigma_{\theta \theta}$ are the radial and circumferential components of the Cauchy stress tensor $\boldsymbol{\sigma}$.
If we model the surrounding material as a neo-Hookean Kelvin-Voigt (NHKV) hyper-viscoelastic material \cite{estrada2018high,mao2017large} where a hyperelastic neo-Hookean spring branch is in parallel with a linear viscosity dash-pot, and assume the surrounding material is incompressible, the stress integral can be calculated analytically as \citep{yang2020extracting} 
\begin{equation}
 S = - \frac{G}{2}  \left[ 5  - \left( \frac{R_{0}}{R} \right)^4  - \frac{4 R_{0}}{R} \right] - \frac{4 \mu \dot{R}}{R}  .
\end{equation} 

Parameter $\mu$ is the viscosity coefficient, and $G$ is the shear modulus. The hyperelastic contribution is described by the strain-energy density function in the form of
\begin{equation}
    W = \frac{G}{2} \left[ (I_1-3) \right],
\end{equation}
\noindent where $I_1 = \text{tr}(\mathbf{C})$ is the first invariant of the right Cauchy-Green deformation tensor $\mathbf{C} = \mathbf{F}^{\top} \mathbf{F}$; $\mathbf{F}$ is the deformation gradient tensor.  %Note that when strain-stiffening is neglected, $\alpha=0$, the stress integral reduces to a special case where the hyperelastic spring branch is a neo-Hookean material model \citep{Gaudron2015bm}, and the overall hyper-viscoelastic material model is called neo-Hookean Kelvin-Voigt (NHKV) model. 
For more complex rheological models, closed-form expressions for the stress integral may not be available, and numerical evaluation is required \cite{wang2026experimental}.

The temporal evolution of the pressure inside the bubble $p_{\rm b}(t)$ is governed by the physics of the bubble contents, which we assume to be a homobaric mixture of water vapor and non-condensable gas\cite{nigmatulin1981dynamics,Akhatov2001hy,barajas2017effects,estrada2018high}. The mathematical description of bubble contents and the resulting evolution equation for $p_{\rm b}(t)$ are summarized in Appendix~\ref{app: bubble contents}. \\

%%%%%%%%%%%%%%%%%%%%%%%%%%%%%%%%%%%%%%%%%%%%
\begin{figure*}[t!]
    \centering
    \includegraphics[width=.8 \textwidth]{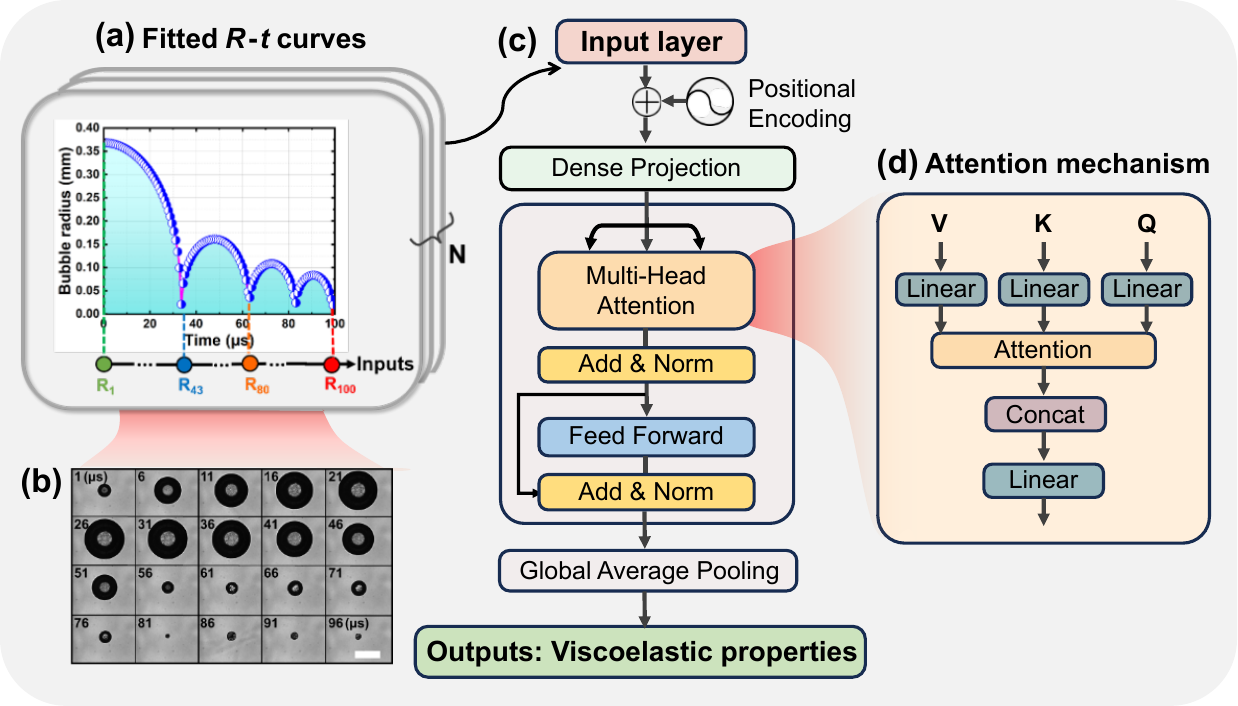}
    \caption{Schematic overview of the BDT model for inferring constitutive properties of soft materials from experimental measurements.
(a) The model input consists of N independent bubble radius versus time ($R$ vs $t$) curves. (b) Selected frames in a typical laser-induced inertial cavitation event in a PAAm 12\% hydrogel (scale bar: 500 $\upmu$m).
(c) Each curve is passed through an input layer with positional encoding, followed by a dense projection, Transformer encoder blocks with multi-head self-attention, and a global average pooling layer.
(d) Within the attention block, the input sequence is transformed into query (\textbf{Q}), key (\textbf{K}), and value (\textbf{V}) matrices to compute temporal dependencies. The model outputs are the inferred viscoelastic parameters: shear modulus ($G$) and viscosity ($\mu$).}
    \label{fig: BDT}
\end{figure*}
%%%%%%%%%%%%%%%%%%%%%%%%%%%%%%%%%%%%%%%%%%%%

\noindent \textbf{Initial and boundary conditions.} 

\vspace{1 em}
Because the initial bubble growth phase of LIC involves laser-induced optical breakdown and plasma formation \cite{liang2022comprehensive}, which are not explicitly resolved in the present cavitation model, numerical simulations are initialized at the experimentally observed maximum bubble radius, where $t$ is set to be 0. Thus, the initial conditions are
\begin{equation}
    R|_{(t=0)}=R_{\max}, \qquad \dot{R}|_{(t=0)}=0 .
\end{equation}

At this instant, the bubble is assumed to be in thermodynamic equilibrium with its surroundings but not in mechanical equilibrium. The subsequent collapse is driven by the pressure imbalance across the bubble interface, including the Laplace pressure contribution. % 
%In this work, we only consider pulsed laser-induced cavitation (LIC). When simulating LIC, we initialize the numerical simulations at the point in time where the bubble has at its maximum radius $(R=R_{\rm max})$. This is done because the initial growth phase of the bubble involves laser breakdown of the liquid and its associated plasma physics, which is not accounted for in the cavitation modeling framework. When the bubble is at its maximum radius, we idealize the system as being in thermodynamic equilibrium with the surroundings, but not in mechanical equilibrium. The driving force for the subsequent cavitation dynamics is the gauge pressure inside the bubble modified by the Laplace pressure $(p_{\rm b}-p_\infty - 2\gamma/R)$, while in LIC, the far-field gauge pressure remains zero $(p_{\rm f}=0)$. For Keller-Miksis-based numerical solutions, the bubble radius and velocity are initialized as $R(t=0)=R_{\rm max}$ and $\dot{R}(t=0) = 0$, respectively.

%%%%%%%%%%%%%%%%%%%%%%%%%%%%%%%%%%%%%%%%%%%%
\begin{figure*}[htbp]
    \centering
    \includegraphics[width=0.8 \textwidth]{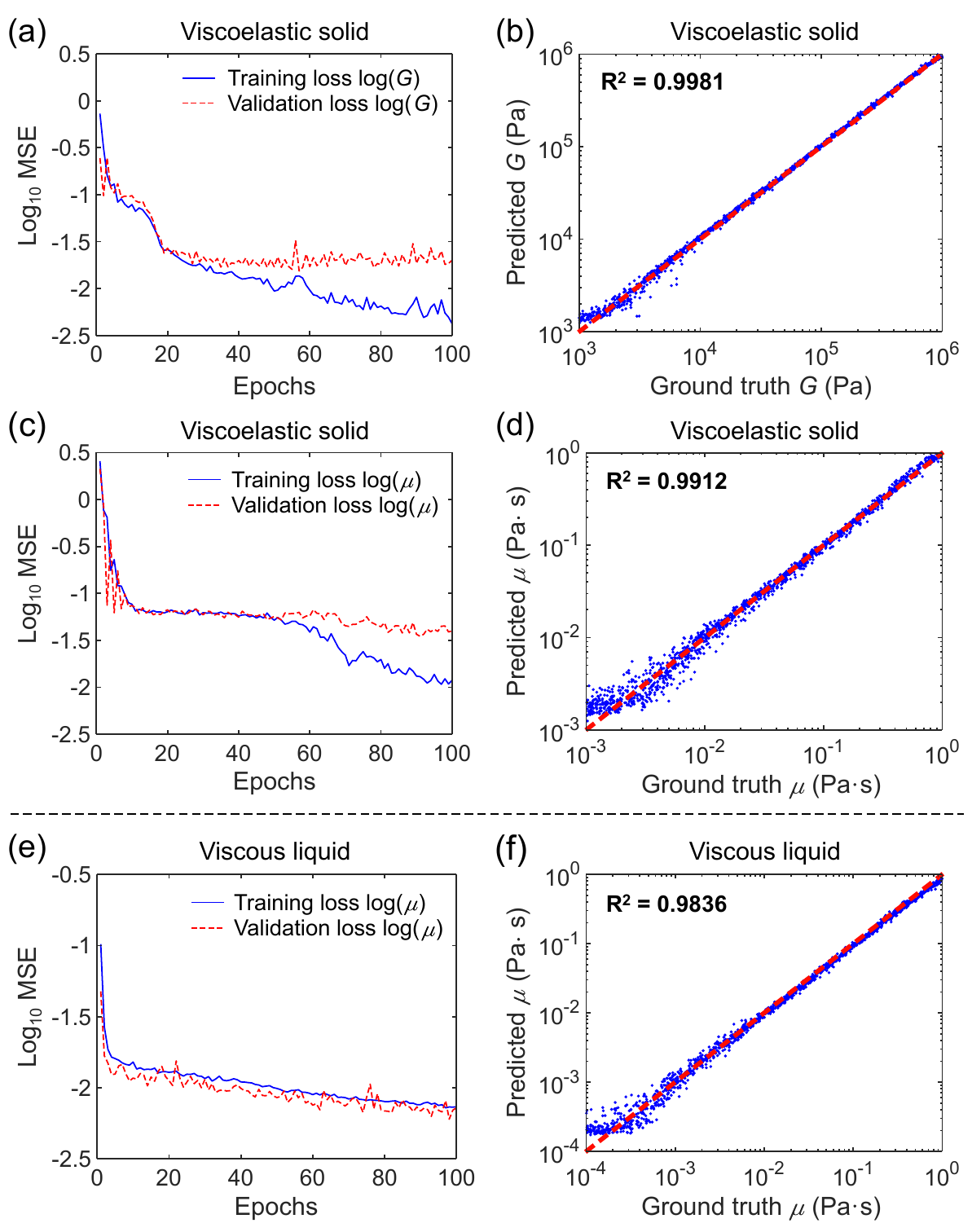}
    \caption{Training and validation performance of the BDT models. (a,c) Training and validation losses for $\log_{10}(G)$ and $\log_{10}(\mu)$ in the BDT-Viscoelasticity model.
(b,d) Predicted versus ground-truth shear modulus $G$ and viscosity $\mu$ for the BDT-Viscoelasticity model, showing strong test-set agreement with $R^2=0.9981$ and $R^2=0.9912$, respectively.
(e) Training and validation losses for $\log_{10}(\mu)$ in the BDT-ViscousOnly model.
(f) Predicted versus ground-truth viscosity $\mu$ for the BDT-ViscousOnly model, with $R^2=0.9836$.
Dashed red lines denote ideal one-to-one agreement.}
    \label{fig: training and validation}
\end{figure*}
%%%%%%%%%%%%%%%%%%%%%%%%%%%%%%%%%%%%%%%%%%%% 

%%%%%%%%%%%%%%%%%%%%%%%%%%%%%%%%%%%%%%%%%%%%
%\begin{figure*}[htbp]
%    \centering
%    \includegraphics[width=0.8\textwidth]{Figures/fig4_relative_error_v3.pdf}
%    \caption{Relative-error distributions of the Bubble Dynamics Transformer (BDT) models on synthetic test datasets.
%(a,b) Error in shear modulus $G$ and viscosity $\mu$ inferred by the BDT-Viscoelasticity model.
%(c) Error in viscosity $\mu$ inferred by the BDT-ViscousOnly model.
%Each blue marker corresponds to one simulated bubble radius--time curve, $R(t)$. The horizontal red dashed line indicates zero relative error. Data points above and below the zero line represent overprediction and underprediction, respectively. The shaded region indicates the error spread.}
%    \label{fig: relative error}
%\end{figure*}
%%%%%%%%%%%%%%%%%%%%%%%%%%%%%%%%%%%%%%%%%%%% 

% ----------------------------------------
\subsection{Bubble Dynamics Transformer Architecture}
\label{sec: BDT}
% ----------------------------------------
%
%
To accelerate inverse constitutive characterization, we developed a Transformer-based deep learning framework, termed the Bubble Dynamics Transformer (BDT), to directly infer viscoelastic material parameters from cavitation bubble dynamics. The BDT maps experimentally measurable bubble radius--time histories, $R(t)$, to constitutive parameters of the surrounding material. In this work, the primary inferred quantities are the shear modulus $G$ and viscosity coefficient $\mu$ in the NHKV model.

The BDT is based on the Transformer architecture \cite{vaswani2017attention}, which uses self-attention mechanisms to capture long-range dependencies in sequential data without recurrent units. This feature is well suited for cavitation-driven inverse mechanics because the material properties are encoded not only in the first collapse but also in the subsequent rebound and damping behavior of the bubble. The multi-head self-attention mechanism enables the model to learn temporal correlations across different stages of the transient bubble response, including collapse, rebound, and relaxation.

Transformer-based models generally include the following core components:  \textit{i) Tokenizers}, which convert text or numerical data into tokens--each token is an integer to represent a character/data or a short segment of characters/data; \textit{ii) An embedding layer}, which converts tokens and positions of tokens into vector representations; \textit{iii) Transformer layers}, which include encoder layers and decoder layers; \textit{iv) An un-embedding layer}, which converts the final vector representations back to a probability distribution over the tokens.\\

\noindent \textbf{i) Tokenizers}. As illustrated in Fig.~\ref{fig: BDT}(a), the input to the BDT model consists of a set of $N$ independent $R$-$t$ curves. Each curve is represented as a sequence of 100 discrete bubble radius data points; for example,  it can be $\lbrace R_1,\ldots,R_{100} \rbrace$ that are sampled at 100 uniformly spaced time points over a 100~$\upmu$s interval with a temporal resolution of 1~$\upmu$s. These curves are collected from different simulations and serve as the foundation for the model to learn from. For simulation datasets whose original temporal discretization does not exactly match this 1~$\upmu$s resolution, each $R$-$t$ curve is resampled onto the prescribed uniform temporal grid using shape-preserving piecewise cubic Hermite interpolation, ensuring that all input curves have an identical sequence length and temporal spacing before tokenization. \\

\noindent \textbf{ii) Embedding layer}. Each token is converted into an embedding vector via a lookup table \cite{vaswani2017attention}. Each of the 100 input radius values is concatenated with a corresponding 64-dimensional positional embedding that encodes its precise location in the temporal sequence \cite{vaswani2017attention}.
To help the model understand not just the size of the bubble at each moment but also when that moment occurs in the sequence, we use a technique called positional encoding. 
Positional encoding is a fixed-size vector representation of the relative positions of tokens within a sequence: it provides the transformer model with information about where the data are in the input sequence. 
This means that each of the 100 radius values is combined with information about its exact position in the timeline. This added information allows the model to distinguish between early and late stages of the collapse, which is crucial for learning the correct patterns.
These embeddings are generated via a learnable embedding layer that maps positional indices to continuous representations \cite{vaswani2017attention}. The combined input—comprising both the raw radius values and their associated positional encodings—is subsequently passed through a nonlinear projection implemented via a dense layer with \textit{softplus} activation\cite{glorot2011deep}, followed by layer normalization (where parameter $\epsilon = 10^{-6}$ serves as a small constant ensuring numerical stability) and dropout regularization (rate = 0.1) to enhance training stability and mitigate overfitting\cite{kowsher2022bangla}.\\

\noindent \textbf{iii) Transformer layers}.  After preparing the input data, we feed it into the heart of the model: the Transformer encoder (see Fig.~\ref{fig: BDT}(c)). Originally developed for language processing (like translating text), the Transformer is very good at recognizing patterns over sequences. It uses a multi-head self-attention mechanism to effectively capture temporal dependencies and long-range interactions inherent in bubble dynamics\cite{liu2021attention, ahmed2023transformers}, which helps the model look at different parts of the time sequence at once and figure out how they relate to each other. For example, it can learn that a sudden decrease in radius is often followed by a rebound, and how these changes relate to the stiffness or viscosity of the material.

The attention mechanism works by turning each time step in the sequence into three mathematical objects—query ($\mathbf{Q}$), key ($\mathbf{K}$), and value ($\mathbf{V}$)—and computing how much attention each part of the sequence should pay to every other part (see Fig.~\ref{fig: BDT}(d)). This helps the model focus on the most important moments in the bubble's behavior.

After the multi-head self-attention step, the encoder employs residual connections and layer normalization (Add \& Norm) to enhance training stability and facilitate efficient gradient propagation\cite{xu2019understanding}. A position-wise feed-forward neural network further refines the encoded features, followed by an additional normalization step to consolidate the learned representations. \\

\noindent  \textbf{iv) Un-embedding layer}.
At the end of the process, the model uses global average pooling to summarize the full sequence into a single, compact set of features. These features are then passed through final layers that produce the predicted microrheology properties, such as the shear modulus ($G$) and viscosity ($\mu$) in the NHKV material model.
% Our direct inference BDT model leverages the learned temporal dependencies within the bubble dynamics, enabling rapid and accurate quantification of material viscoelasticity.

%%%%%%%%%%%%%%%%%%%%%%%%%%%%%%%%%%%%%%%%%%%%
\begin{figure*}[htbp]
    \centering
    \includegraphics[width=1 \textwidth]{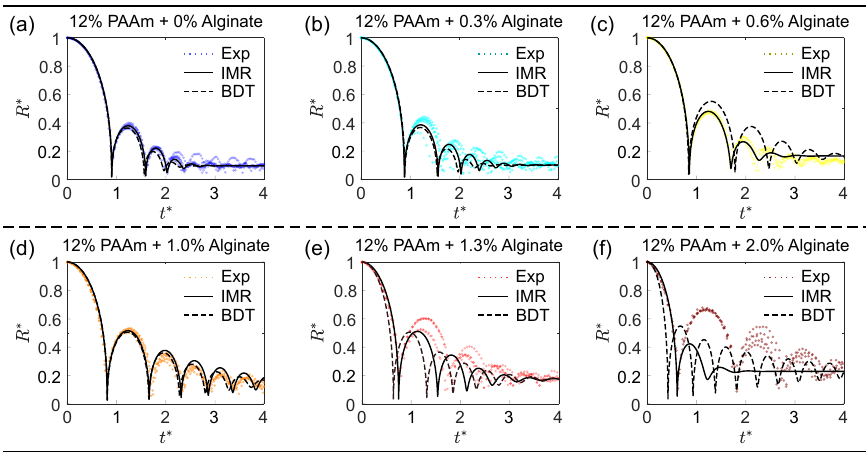}
    \caption{Experimental validation of the BDT-Viscoelasticity model for PAAm-alginate double-network hydrogels.
(a--f) Normalized bubble radius dynamics, $R^*$ versus $t^*$, for 12\% PAAm hydrogels with alginate concentrations from 0\% to 2.0\%. Dotted curves denote experimental measurements, solid curves denote conventional IMR fits, and dashed curves denote BDT inferences. The BDT model reproduces the main collapse and rebound features of the measured cavitation dynamics across different stiffness and damping regimes when the alginate concentration is smaller than 1.0\%. The inferred material parameters are summarized in Table~\ref{tab: extracted material_properties_PAAm}.}
\label{fig: PAAm_Alginate_inference_results}
\end{figure*}
%%%%%%%%%%%%%%%%%%%%%%%%%%%%%%%%%%%%%%%%%%%%

%%%%%%%%%%%%%%%%%%%%%%%%%%%%%%%%%%%%%%%%%%%%
\begin{figure*}[htbp]
    \centering
    \includegraphics[width=1 \textwidth]{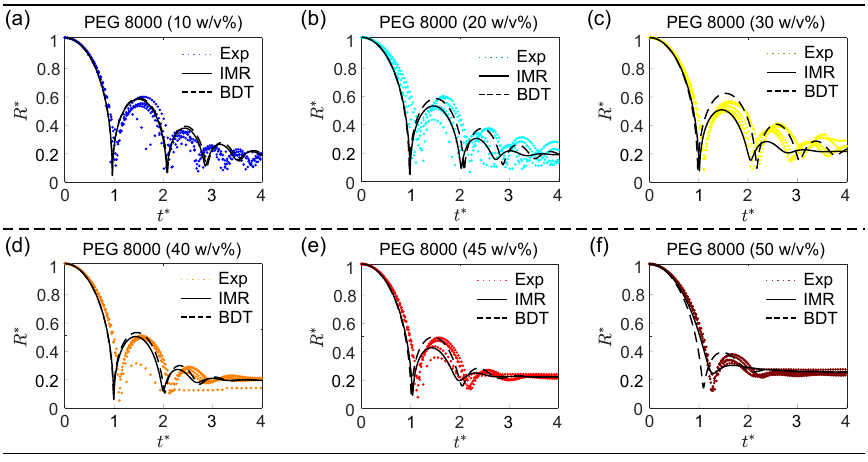}
    \caption{Experimental validation of the BDT-ViscousOnly model for PEG 8000/water solutions.
(a--f) Normalized bubble radius dynamics, $R^*$ versus $t^*$, for PEG 8000 concentrations from 10 to 50~w/v\%.
Dotted curves indicate experimental measurements, solid curves indicate conventional IMR fits, and dashed curves indicate BDT-ViscousOnly inferences. The model reproduces the main collapse, rebound, and damping features of the measured cavitation dynamics. The inferred viscosity parameters are summarized in Table~\ref{tab: extracted material_properties_PEG8K}.}
    \label{fig: PEG8K_inference_results}
\end{figure*}
%%%%%%%%%%%%%%%%%%%%%%%%%%%%%%%%%%%%%%%%%%%%

% =================================
\subsection{Synthetic training data generation}
\label{sec:training_data}
% =================================
The BDT models were trained using synthetic datasets generated from a previously developed IMR numerical simulation framework \cite{tzoumaka2023modeling}. The simulations used the NHKV constitutive model, in which the surrounding material is represented by a neo-Hookean hyperelastic spring in parallel with a linear viscous dashpot. 

For the BDT-Viscoelasticity model, which infers both $G$ and $\mu$, the shear modulus and viscosity were sampled from uniform distributions in logarithmic space: specifically, $\log_{10}(G) \in [3, 6]$ (unit: Pa) and $\log_{10}(\mu) \in [-3, 0]$ (unit: Pa$\cdot$s). To introduce variability in the collapse dynamics, the maximum bubble expansion ratio, $\lambda = R_{\max}/R_0$, was also sampled uniformly at random from the interval $[3, 11]$.

For the BDT-ViscousOnly model, which infers viscosity for viscous-liquid-like materials, the elastic contribution was disabled by setting $G=0$. The viscosity was sampled from
  $\log_{10} \mu \in [-4,0]$ (Pa$\cdot$s), and $\lambda$ was likewise sampled uniformly from  $[3, 11]$.
These ranges were chosen based on our experience from previously tested soft materials and from data observed in a wide range of soft materials \cite{estrada2018high,yang2019strain,yang2020extracting,yang2021predicting,yang2022mechanical,mancia2021acoustic,tzoumaka2023modeling,mcghee2023high,yang2024estimating,bremer2024ballistic} from soft hydrogel/tissue to relatively stiffer elastomers.

Using these parameter ranges, 14{,}160 distinct bubble-dynamics simulations were generated for the BDT-Viscoelasticity training dataset, with each curve corresponding to an independently sampled parameter set \(\{G,\mu,\lambda\}\). For the BDT-ViscousOnly model,  {15{,}416} simulations were generated. Each simulation produced a bubble radius--time curve sampled at 100 uniformly spaced time points over a 100~$\upmu$s interval (temporal resolution of 1~$\upmu$s). The input to the BDT model was the sampled $R(t)$ curves, and the target outputs were $\{\log_{10}G,\log_{10}\mu\}$ for BDT-Viscoelasticity and $\log_{10}\mu$ for BDT-ViscousOnly.

% =================================
\subsection{Training and validation}

Both BDT models were trained using the Adam optimizer with a learning rate of 0.001 and a batch size of 32. Both models were trained for a fixed budget of 100 epochs to ensure sufficient convergence and maintain a consistent training protocol. The validation losses stabilized within this training window without substantial deterioration at later epochs. The network used \textit{softplus} activation functions, residual connections, dropout regularization, and layer normalization with $\varepsilon=10^{-6}$ to improve numerical stability and reduce overfitting.

The loss function was the mean squared error (MSE) between the inferred and ground-truth logarithmic material parameters:
\begin{equation}
\text{MSE} = \frac{1}{N} \sum_{i=1}^{N} \left( \hat{y}_i - y_i \right)^2,
\label{eq:mse}
\end{equation}
where $N$ is the number of samples, $y_i$ is the ground truth value of the $i$-th data, and $\hat{y}_i$ is the corresponding model prediction.

To ensure effective model training, 80\% of the dataset was used for training, with 10\% allocated for validation. The validation set was used to monitor model convergence and assess potential overfitting during training, while the independent test set was used only for final performance evaluation. As shown in Fig.~\ref{fig: training and validation}(a) and (c), the BDT-Viscoelasticity model demonstrated convergence for both $\log(G)$ and $\log(\mu)$ within the prescribed maximum of 100 epochs. The checkpoint with the lowest validation loss was subsequently used for test-set evaluation. The remaining 10\% of the dataset was reserved for testing the model's predictive performance by comparing the predicted values of $G$ and $\mu$ against their corresponding ground truth values. The coefficient of determination ($R^2$) was used as an evaluation metric\cite{allen2022machine}. As illustrated in Fig.~\ref{fig: training and validation}(b) and (d), the \(R^2\) values for the predicted shear modulus (\(G\)) and viscosity (\(\mu\)) are 0.9981 and 0.9912, respectively, indicating excellent agreement with the ground truth and demonstrating the high accuracy of the model in capturing material properties. As shown in Fig.~\ref{fig: training and validation}(e), the training and validation losses of the BDT-ViscousOnly model for $\log(\mu)$ converge within the 100-epoch training budget, and the checkpoint with the lowest validation loss was used for subsequent evaluation; Fig.~\ref{fig: training and validation}(f) shows excellent inference-ground-truth agreement on the synthetic test set ($R^2=0.9836$).

%To examine both the magnitude and direction of the inference errors, Fig.~\ref{fig: relative error} reports the \emph{relative error} as a function of the ground-truth material parameter on the synthetic test set. Positive and negative values indicate overprediction and underprediction, respectively, while the horizontal red line indicates zero error.  Specifically, or each sample we define the relative error as
%\begin{equation}
%\mathrm{Relative\ Error}~(\%) = \frac{\hat{y}-y}{y}\times 100\%,
%\end{equation}
%where $y$ denotes the ground-truth value ($\it{G}$ or $\it{\mu}$) and $\hat{y}$ is the corresponding model inference. 

%For the BDT-Viscoelasticity, the shear-modulus predictions remain tightly centered about zero error across the sampled $\it{G}$ range (Fig.~\ref{fig: relative error}a), whereas viscosity predictions exhibit a broader, heteroscedastic error distribution that increases toward smaller $\it{\mu}$ (Fig.~\ref{fig: relative error}b). A similar trend is observed for the BDT-ViscousOnly model evaluated over its expanded viscosity range (Fig.~\ref{fig: relative error}(c), underscoring that viscosity inference from bubble dynamics is intrinsically more challenging, especially in the low-$\it{\mu}$ regime.

%!%!%!%!%!%!%!%!%!%!%!%!%!%!%!%!%%!%!%!%!%!%!%!%!%%!%!%!
\begin{table*}[htbp]
\caption{Estimated viscoelastic material properties for PAAm alginate hydrogels} \label{tab: extracted material_properties_PAAm}
%\normalsize
\begin{ruledtabular}
\begin{tabular}{cccccc}
Alginate \%            & Method & $G$ (kPa)                       & $\mu$ (Pa$\cdot$s)                        & Relative error of $G$ & Relative error of $\mu$ \\ \hline
\multirow{3}{*}{0\%}   & LIC test: IMR    & 4.20 $\pm$ 1.21  & 0.042 $\pm$ 0.008 & --              & --                \\ 
                       & LIC test: BDT    & 4.83 $\pm$ 0.81  & 0.055 $\pm$ 0.008 &   15.0 \%               & 32.3 \%
\\ 
                       & Rheometer test    & 1.10 $\pm$ 0.38  & 40.84 $\pm$ 5.84 &   --                & -- 
                       
                       \\ 
                       \arrayrulecolor{lightgray}\hline
\multirow{3}{*}{0.3\%} & LIC test: IMR    & 6.44 $\pm$ 1.64  & 0.020 $\pm$ 0.009 & --               & --                \\
                       & LIC test: BDT    & 5.94 $\pm$ 1.03                  & 0.035 $\pm$ 0.010                 &     -7.8\%              &     70.0\%   \\ 
                       & Rheometer test    & 1.30 $\pm$ 0.07  & 42.62 $\pm$ 6.07 &   --                & -- 
                       
                       \\ \arrayrulecolor{lightgray}\hline
\multirow{3}{*}{0.6\%} & LIC test: IMR    & 13.82 $\pm$ 0.91 & 0.095 $\pm$ 0.019                 & --               & --                \\
                       & LIC test: BDT    & 13.29 $\pm$ 1.72                 & 0.041 $\pm$ 0.008                 &     -3.8\%              &    -56.6\%
                        \\ 
                       & Rheometer test    & 2.19 $\pm$ 0.04 & 83.19 $\pm$ 11.22 &   --                 & -- 
                       \\ \hline
\multirow{3}{*}{1.0\%} & LIC test: IMR    & 15.71 $\pm$ 1.18                 & 0.008 $\pm$ 0.004                 & --               & --                \\
                       & LIC test: BDT    & 15.43 $\pm$ 1.13                 & 0.014 $\pm$ 0.003                 &    -1.8\%              &    87.0\%    \\ 
                       & Rheometer test    & 3.50 $\pm$ 0.18 & 171.93 $\pm$ 23.14 &   --                 & --             
                       \\ \hline
\multirow{3}{*}{1.3\%} & LIC test: IMR    & 29.71 $\pm$ 3.49                 & 0.036 $\pm$ 0.044                 & --               & --                \\
                       & LIC test: BDT    & 57.58 $\pm$ 1.02                 & 0.025 $\pm$ 0.002                 &    93.8\%              &    -32.1\%   \\ 
                       & Rheometer test    & 5.81 $\pm$ 0.08  & 311.85 $\pm$ 33.74 &   --                & --               
                       \\ \hline
\multirow{3}{*}{2.0\%} & LIC test: IMR    & 85.81 $\pm$ 5.43                 & 0.128 $\pm$ 0.048                 & --               & --                \\
                       & LIC test: BDT    & 203.6 $\pm$ 36.8               & 0.004 $\pm$ 0.001                 &  137.2\%                 &   -96.8\%  \\ 
                       & Rheometer test    & 10.95 $\pm$ 0.45   & 841.8 $\pm$ 166.5 &  --                & --            
                       
                       \\ \arrayrulecolor{black}
\end{tabular}
\end{ruledtabular}
\end{table*}
%!%!%!%!%!%!%!%!%!%!%!%!%!%!%!%!%%!%!%!%!%!%!%!%!%%!%!%!

%!%!%!%!%!%!%!%!%!%!%!%!%!%!%!%!%%!%!%!%!%!%!%!%!%%!%!%!
\begin{table*}[htbp]
\caption{Estimated viscosity material properties for PEG 8000/water mixtures} \label{tab: extracted material_properties_PEG8K}
%\normalsize
\begin{ruledtabular}
\begin{tabular}{cccccc}
w/v\%            & Method                       & $\mu$ (Pa$\cdot$s)                         & Relative error of $\mu$ \\ \hline
\multirow{3}{*}{10\%}   & LIC test: IMR     & 0.047 $\pm$ 0.016 & --                              \\ 
                       & LIC test: BDT      & 0.043 $\pm$ 0.008               & -8.65 \%
\\ 
                       & Rheometer test     & 0.0064 $\pm$ 0.0004                & -- 
                       
                       \\ 
                       \arrayrulecolor{lightgray}\hline

\multirow{3}{*}{20\%}   & LIC test: IMR     & 0.083 $\pm$ 0.023 & --                               \\ 
                       & LIC test: BDT      & 0.055 $\pm$ 0.017              & -33.62 \%
\\ 
                       & Rheometer test     & 0.015 $\pm$ 0.007 & --               
                       
                       \\ 
                       \arrayrulecolor{lightgray}\hline

\multirow{3}{*}{30\%}   & LIC test: IMR     & 0.167 $\pm$ 0.024 & --                              \\ 
                       & LIC test: BDT      & 0.085 $\pm$ 0.012               & -49.07 \%
\\ 
                       & Rheometer test     & 0.037 $\pm$ 0.020               & -- 
                       
                       \\ 
                       \arrayrulecolor{lightgray}\hline

\multirow{3}{*}{40\%}   & LIC test: IMR     & 0.111 $\pm$ 0.023 & --                               \\ 
                       & LIC test: BDT      & 0.095 $\pm$ 0.024                & -14.36 \%
\\ 
                       & Rheometer test     & 0.130 $\pm$ 0.012               & -- 
                       
                       \\ 
                       \arrayrulecolor{lightgray}\hline

\multirow{3}{*}{45\%}   & LIC test: IMR     & 0.185 $\pm$ 0.020 & --                               \\ 
                       & LIC test: BDT      & 0.136 $\pm$ 0.016               & -26.80 \%
\\ 
                       & Rheometer test     &  76.15 $\pm$ 17.30                & -- 
                       
                       \\ 
                       \arrayrulecolor{lightgray}\hline

\multirow{3}{*}{50\%}   & LIC test: IMR     & 0.533 $\pm$ 0.097 & --                               \\ 
                       & LIC test: BDT      & 0.277 $\pm$ 0.027               & -48.07 \%
\\ 
                       & Rheometer test     & 105.10 $\pm$ 34.21              & --

                       \\ \arrayrulecolor{black}
\end{tabular}
\end{ruledtabular}
\end{table*}
%!%!%!%!%!%!%!%!%!%!%!%!%!%!%!%!%%!%!%!%!%!%!%!%!%%!%!%!

%!%!%!%!%!%!%!%!%!%!%!%!%!%!%!%!%%!%!%!%!%!%!%!%!%%!%!%!
\begin{table*}[htbp]
\caption{Estimated viscoelastic material properties of different hydrogel systems from LIC experiments using IMR and BDT. The IMR values are taken from the previously reported study by Yang et al.\cite{yang2024estimating}} \label{tab: extracted_material_properties_hydogels}
\begin{ruledtabular}
\begin{tabular}{cccccc}
Material & Method & $G$ (kPa) & $\mu$ (Pa$\cdot$s) & Relative error of $G$ & Relative error of $\mu$ \\ \hline

\multirow{2}{*}{PAAm 3\%}
& LIC test: IMR & 8.31 $\pm$ 0.43 & 0.093 $\pm$ 0.073 & -- & -- \\
& LIC test: BDT & 2.94 $\pm$ 0.35 & 0.156 $\pm$ 0.239 & -64.6\% & 67.7\% \\
\arrayrulecolor{lightgray}\hline

\multirow{2}{*}{PAAm 8\%}
& LIC test: IMR & 15.1 $\pm$ 4.4 & 0.209 $\pm$ 0.180 & -- & -- \\
& LIC test: BDT & 22.6 $\pm$ 9.9 & 0.079 $\pm$ 0.017 & 49.8\% & -62.3\% \\
\arrayrulecolor{lightgray}\hline

\multirow{2}{*}{Agarose 0.5\%}
& LIC test: IMR & 27.8 $\pm$ 8.6 & 0.17 $\pm$ 0.09 & -- & -- \\
& LIC test: BDT & 27.0 $\pm$ 24.4 & 0.018 $\pm$ 0.013 & -2.9\% & -89.2\% \\
\arrayrulecolor{lightgray}\hline

\multirow{2}{*}{Agarose 1.0\%}
& LIC test: IMR & 58.1 $\pm$ 9.4 & 0.30 $\pm$ 0.09 & -- & -- \\
& LIC test: BDT & 64.8 $\pm$ 12.4 & 0.023 $\pm$ 0.005 & 11.4\% & -92.3\% \\
\arrayrulecolor{lightgray}\hline

\multirow{2}{*}{Agarose 2.5\%}
& LIC test: IMR & 114.6 $\pm$ 4.9 & 0.11 $\pm$ 0.05 & -- & -- \\
& LIC test: BDT & 155.6 $\pm$ 11.7 & 0.021 $\pm$ 0.005 & 35.7\% & -80.6\% \\
\arrayrulecolor{lightgray}\hline

\multirow{2}{*}{Agarose 5.0\%}
& LIC test: IMR & 282.4 $\pm$ 22.8 & 0.50 $\pm$ 0.07 & -- & -- \\
& LIC test: BDT & 344.8 $\pm$ 15.6 & 0.034 $\pm$ 0.008 & 22.1\% & -93.3\% \\
\arrayrulecolor{lightgray}\hline

\multirow{2}{*}{Gelatin 6\%}
& LIC test: IMR & 12.9 $\pm$ 3.7 & 0.027 $\pm$ 0.022 & -- & -- \\
& LIC test: BDT & 12.7 $\pm$ 5.9 & 0.017 $\pm$ 0.004 & -1.0\% & -37.8\% \\
\arrayrulecolor{lightgray}\hline

\multirow{2}{*}{Gelatin 10\%}
& LIC test: IMR & 29.9 $\pm$ 6.2 & 0.055 $\pm$ 0.061 & -- & -- \\
& LIC test: BDT & 39.1 $\pm$ 12.8 & 0.021 $\pm$ 0.015 & 30.8\% & -62.7\% \\
\arrayrulecolor{lightgray}\hline

\multirow{2}{*}{Gelatin 14\%}
& LIC test: IMR & 101.5 $\pm$ 4.0 & 0.166 $\pm$ 0.208 & -- & -- \\
& LIC test: BDT & 147.7 $\pm$ 13.2 & 0.081 $\pm$ 0.002 & 45.5\% & -51.0\% \\

\arrayrulecolor{black}
\end{tabular}
\end{ruledtabular}
\end{table*}
%!%!%!%!%!%!%!%!%!%!%!%!%!%!%!%!%%!%!%!%!%!%!%!%!%%!%!%!

% =================================
\section{Results}
\label{sec:results}
% =================================
To assess the inference capability of the developed BDT model, we first applied the BDT-Viscoelasticity model to experimental LIC data obtained from PAAm-alginate hydrogels. The hydrogels were prepared with alginate concentrations of 0\%, 0.3\%, 0.6\%, 1.0\%, 1.3\%, and 2.0\%, as described in Section~\ref{sec:exp}.
Figure~\ref{fig: PAAm_Alginate_inference_results} displays the normalized bubble radius dynamics ($R^{\star}$) as a function of non-dimensional time ($t^{\star}$) for six representative hydrogel formulations.  Each subfigure in Fig.~\ref{fig: PAAm_Alginate_inference_results} compares the experimental measurements (dotted lines) with inference from the conventional IMR model (solid lines) and our proposed BDT-Viscoelasticity model (dashed lines). Across all tested alginate concentrations, the BDT-Viscoelasticity model captures the main collapse of the measured cavitation dynamics for alginate concentrations from 0\% to 1.0\%, where the BDT-inferred dynamics show close agreement with the experimental trajectories and are comparable to the conventional IMR fits.   

A quantitative comparison of the shear modulus $G$ and viscosity $\mu$ inferred by BDT-Viscoelasticity and conventional IMR is summarized in Table~\ref{tab: extracted material_properties_PAAm}. Rheometer measurements are also included as low-frequency linear viscoelastic benchmarks. The BDT-Viscoelasticity model accurately infers the shear modulus for hydrogels with alginate concentrations up to 1.0\%, with relative errors below 15\% compared with conventional IMR. Larger discrepancies are observed for the 1.3\% and 2.0\% alginate samples, where the relative errors in $G$ increase to 93.8\% and 137.2\%, respectively. These deviations are likely associated with stronger nonlinear stiffening and the possible emergence of additional mechanical mechanisms, such as damage, fracture, or irreversible network rearrangement, that are not fully represented by the NHKV model used to generate the synthetic training data.

Comparison between the LIC-based high-rate estimates and the rheometer measurements indicates pronounced rate-dependent stiffening of the PAAm-alginate hydrogels. For all formulations, the shear modulus inferred from LIC is substantially higher than the low-frequency rheometer value. This difference becomes more pronounced with increasing alginate concentration, suggesting that alginate-rich double-network hydrogels exhibit stronger stiffening under the ultra-high strain-rate deformation generated by inertial cavitation.

The inferred viscosity coefficient $\mu$ generally exhibits larger errors than the inferred shear modulus $G$. This trend is consistent with previous IMR analyses \cite{tzoumaka2023modeling} and reflects the reduced sensitivity of the first collapse regime of the bubble radius--time response to viscous effects compared with elastic stiffness. Consequently, viscosity inference from cavitation dynamics remains more challenging than shear-modulus inference, particularly when the dissipative contribution is weak or when the material response deviates from the assumed NHKV form.

To further contextualize the composition-dependent rheological response of PAAm alginate hydrogels, Fig.~\ref{fig: PAAm_Alginate_rheometer_results} presents independent rheometer characterization. The storage modulus $G'$ exhibits a low-strain plateau over the linear viscoelastic regime, followed by progressive softening at larger oscillatory strains, with the onset and magnitude of nonlinearity dependent on alginate content. The loss modulus $G''$ shows an increasing dissipative response with strain, consistent with enhanced energy dissipation and network rearrangement under large deformation.
For rheometer measurements, the reported viscosity was calculated as 
$\mu_{\mathrm{rheo}}=G''/\omega$ within the linear viscoelastic regime. 
Although this definition is consistent with the dashpot coefficient of a 
linear Kelvin--Voigt model under small-amplitude oscillatory shear, it 
should not be interpreted as directly equivalent to the IMR/BDT-inferred 
$\mu$. The latter represents an effective high-rate damping parameter 
identified from transient cavitation dynamics involving finite deformation,  spatially nonuniform strain rates, and microsecond-scale collapse and rebound.
Together, these rheometer trends provide quasi-static benchmarks confirming the composition-dependent stiffness and damping of the hydrogels, while emphasizing that the LIC-based IMR and BDT inference probes a distinct ultra-high-rate deformation regime where viscoelastic parameters can differ substantially from low-frequency measurements.

We next evaluated the BDT-ViscousOnly model using LIC experiments performed on PEG 8000/water mixtures with PEG 8000 concentrations ranging from 10 to 50~w/v\%. Figure~\ref{fig: PEG8K_inference_results} compares the normalized experimental bubble dynamics with conventional IMR fits and BDT-ViscousOnly inferences. Across the tested concentration range, the BDT-ViscousOnly model reproduces the primary collapse, rebound, and damping features of the measured radius--time trajectories. These results demonstrate that a reduced inference model, in which the elastic contribution is neglected, can recover effective viscous dissipation behavior from LIC dynamics in viscous polymer solutions.

The inferred viscosity coefficients are summarized in Table~\ref{tab: extracted material_properties_PEG8K} and compared with conventional IMR estimates. Across the tested PEG 8000/water mixtures, the IMR-inferred viscosities span approximately $0.05$--$0.53~\mathrm{Pa\cdot s}$. The BDT-ViscousOnly model yields viscosity values of comparable magnitude to those obtained from IMR, demonstrating its ability to recover the effective dissipative response governing the measured cavitation dynamics. Complementary oscillatory shear rheometer measurements for the PEG 8000/water mixtures are shown in Fig.~\ref{fig: PEG8K_rheometer_results}. The frequency-sweep results reveal concentration-dependent variations in the loss modulus and the corresponding viscosity coefficient, $\mu_{\mathrm{rheo}}=G''/\omega$, measured within the low-frequency linear viscoelastic regime. The discrepancy between the rheometer-derived viscosity and the IMR/BDT-inferred viscosity arises because the two measurements probe fundamentally different deformation regimes. Rheometer measurements provide a low-frequency, small-amplitude reference response, whereas LIC-based IMR and BDT infer an effective high-rate viscosity associated with transient cavitation dynamics involving localized, finite-deformation, and ultra-high strain-rate loading.

The results summarized in Table~\ref{tab: extracted_material_properties_hydogels} further evaluate the transferability of the BDT-Viscoelasticity model to previously reported LIC datasets for different hydrogel systems. Overall, the agreement between BDT and conventional IMR depends strongly on how closely the material response conforms to the NHKV constitutive assumptions used to generate the synthetic training data. For PAAm 3\% and 8\%, relatively large discrepancies are observed, suggesting that their cavitation responses may involve mechanical features not fully represented by the NHKV training model. In contrast, low-concentration agarose samples, particularly 0.5\% and 1.0\% agarose, show better agreement in the inferred shear modulus, indicating that their dynamic response remains closer to the constitutive behavior represented in the training dataset. Similar agreement is observed for 6\% gelatin.

For higher-concentration agarose and gelatin samples, the discrepancies between BDT and IMR generally increase, especially in the inferred viscosity coefficient. These deviations may arise from stronger nonlinear elasticity, rate-dependent stiffening, heterogeneous dissipation, or additional relaxation mechanisms that are not captured by the simplified NHKV model. The viscosity inference is particularly sensitive to such deviations because the bubble radius--time response is less sensitive to viscous damping than to elastic stiffness during the first collapse. Therefore, the results in Table~\ref{tab: extracted_material_properties_hydogels} indicate that the present BDT model performs best when the target material response remains within the constitutive space spanned by the NHKV-based synthetic training data. Extending the training dataset to include broader constitutive models and additional dissipative mechanisms will be important for improving BDT generalizability across a wider range of soft materials.

\subsection*{Computational Cost} 
The BDT-Viscoelasticity and BDT-ViscousOnly models were implemented in Python 3.9.18 using TensorFlow 2.7.0 and trained on a workstation equipped with an NVIDIA GeForce RTX~4070~Ti GPU.

Synthetic dataset generation was performed in \textsc{Matlab} R2023b on a Windows~10 workstation equipped with an Intel Core i7-13700KF CPU, 64~GB of memory, and 24 computational threads. The datasets contained 14{,}160 and 15{,}416 simulated bubble radius--time curves, $R(t)$, for the BDT-Viscoelasticity and BDT-ViscousOnly models, respectively. Dataset generation was parallelized using the \textsc{Matlab} Parallel Computing Toolbox and required approximately two weeks.

The BDT-Viscoelasticity model was trained in approximately 10~min. Once trained, the model enabled rapid inference of material properties from experimental cavitation trajectories. For the PAAm-alginate hydrogels with alginate concentrations ranging from 0\% to 2.0\%, the average inference times were 0.613, 0.658, 0.617, 0.606, 0.616, and 0.639~s, respectively. The BDT-ViscousOnly model was trained in approximately 8~min. After training, it enabled rapid inference of viscosity from experimental bubble dynamics. For PEG 8000/water mixtures with PEG concentrations ranging from 10 to 50~w/v\%, the average inference times were 0.712, 0.768, 0.689, 0.643, 0.651, and 0.684~s, respectively. These results demonstrate that, after the offline cost of physics-based dataset generation and model training, the trained BDT models provide rapid material-parameter inference without requiring repeated iterative forward-model fitting.

\subsection*{Online Platform of the Bubble Dynamics Transformer}

To facilitate reproducible use and broader dissemination of the proposed framework, we developed browser-based inference platforms for the BDT models and deployed them as public Hugging Face Spaces. The BDT-Viscoelasticity platform is available at \url{https://huggingface.co/spaces/lehuaaa/BDT}. The interface provides a guided end-to-end workflow comprising (i) experimental data upload and parsing, (ii) preprocessing and interpolation of the bubble radius--time trajectory, (iii) machine-learning inference of material parameters, (iv) physics-based forward simulation for consistency checking, and (v) export of predicted properties and reconstructed dynamics. Given an experimental $R(t)$ curve, the platform returns estimates of the shear modulus $G$ and viscosity $\mu$ and enables quantitative comparison between measurements and model reconstructions generated using the inferred parameters. 

A separate BDT-ViscousOnly platform is provided for applications in which only the viscosity coefficient $\mu$ is inferred: \url{https://huggingface.co/spaces/lehuaaa/BDT\_Viscosity\_Only}. The associated user guide, training scripts, pretrained models, and sample test data are openly available at \url{https://github.com/YangMechanicsGroupUTAustin/Bubble_Dynamics_Transformer_Model}. These resources are intended to support reproducibility, rapid evaluation, and broader community adoption of the proposed AI-enhanced inverse microrheology framework.

% =================================
\section{Conclusions}
\label{sec:conclusions}
% =================================
This work developed an AI-enhanced inverse microrheology framework to accelerate material characterization at ultra-high strain rates. Recent IMR provides a powerful approach for extracting soft material properties at strain rates exceeding $10^{3}$~s$^{-1}$ by analyzing transient inertial cavitation dynamics. However, conventional IMR requires iterative physics-based forward-model fitting, which can be computationally expensive and may limit its use in high-throughput or near-real-time applications. To address this limitation, we introduced the Bubble Dynamics Transformer (BDT), a Transformer-based deep learning framework designed to infer viscoelastic and viscous material parameters directly from experimentally measured bubble radius--time trajectories.

The BDT models demonstrated robust training behavior and accurate inference across a broad range of stiffness and damping conditions represented in the synthetic training datasets. The BDT-Viscoelasticity model inferred both shear modulus $G$ and viscosity coefficient $\mu$ from normalized LIC bubble dynamics and achieved close agreement with conventional IMR estimates for several hydrogel systems. The BDT-ViscousOnly model further enabled rapid viscosity inference for viscous polymer solutions, providing a streamlined framework for materials whose cavitation response is dominated by viscous dissipation. Once trained, both models enabled material-parameter inference on the order of seconds, thereby substantially reducing the computational cost associated with repeated iterative IMR fitting.

The results also highlight an important limitation of the present framework. Because the current BDT models were trained using synthetic data generated from the NHKV constitutive model, their inference accuracy depends on how closely the target material response conforms to this constitutive assumption. For materials exhibiting stronger nonlinear stiffening, damage, fracture, plasticity, heterogeneous dissipation, or additional relaxation mechanisms during cavitation, both conventional IMR models based on simplified constitutive laws and the current BDT models may exhibit reduced accuracy. Future work should therefore incorporate broader constitutive descriptions, damage and failure mechanisms, and richer dissipative physics into the bubble-dynamics simulations used for BDT training.

In summary, the proposed BDT framework establishes a data-driven and physics-informed route for rapid inverse characterization of soft materials from transient cavitation dynamics. By combining physics-based synthetic training data, high-speed experimental measurements, and Transformer-based sequence learning, this approach provides a new pathway for AI-enhanced experimental mechanics and intelligent constitutive characterization under extreme loading conditions. Future developments may include uncertainty quantification, active learning for adaptive dataset generation, expanded constitutive model libraries, and real-time deployment for high-throughput material screening, soft tissue characterization, hydrogel design, and biomedical device evaluation.

% =================================
\section*{Acknowledgment}
We gratefully acknowledge support by the U.S. National Science Foundation (NSF) under Grant No. 2232428 and 2441460. J.Y. acknowledges the start-up fund from the University of Texas at Austin, the Haythornthwaite Foundation for the Research Initiation Grant awarded by the Applied Mechanics Division of the American Society of Mechanical Engineers (ASME).
L.B. would like to acknowledge the Graduate Excellence Fellowship and Professional Development Award from the University of Texas at Austin Cockrell School of Engineering. 
This material is based upon work partially supported by the U.S. National Science Foundation under award No. 2452029 (J.F. and J.Y.).  
The opinions, findings, and conclusions, or recommendations expressed are those of the authors and do not necessarily reflect the views of the NSF.

% =================================
%\section*{{Compliance with Ethical Standards}}
% \textbf{Conflict of interest} The authors declare that they have no conflict of interest. 

\appendix

\section{Oscillatory Shear Plate  Rheometer Results}
\label{app: rheometer results}

\begin{figure*}%[h]
    \centering
    \includegraphics[width=1 \textwidth]{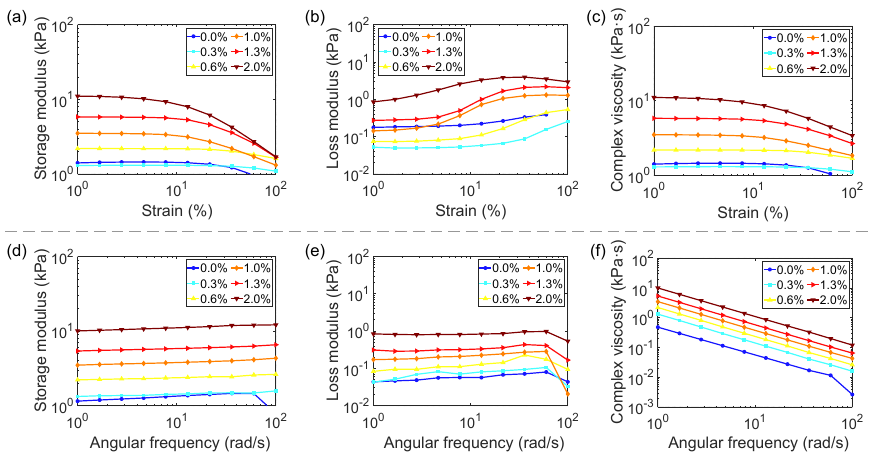}
  \caption{Oscillatory shear rheometry of ``12\% PAAm + X\% alginate'' double-network hydrogels.
(a--c) Strain-dependent storage modulus $G'$, loss modulus $G''$, and complex viscosity $|\eta^*|$.
(d--f) Angular-frequency-dependent $G'$, $G''$, and $|\eta^*|$ measured within the linear viscoelastic regime.
Increasing alginate concentration increases both elastic stiffness and dissipative response.} %, whereas the complex viscosity decreases with increasing angular frequency for all samples.}
    \label{fig: PAAm_Alginate_rheometer_results}
\end{figure*}

\begin{figure*}%[h]
    \centering
    \includegraphics[width=1 \textwidth]{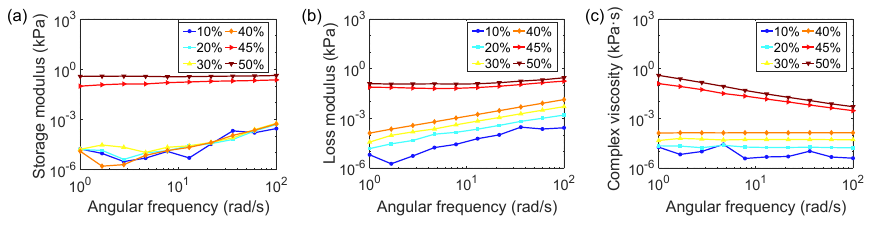}
\caption{Oscillatory shear rheometry of PEG 8000/water mixtures.
(a) Storage modulus $G'$, (b) loss modulus $G''$, and (c) complex viscosity $|\eta^*|$ as functions of angular frequency for PEG 8000 concentrations of 10--50~w/v\%.
Increasing PEG concentration enhances the viscoelastic response, while the complex viscosity decreases with increasing angular frequency.}
    \label{fig: PEG8K_rheometer_results}
\end{figure*}

Rheometer tests were performed using an oscillatory shear plate rheometer. Hydrogel samples were tested with an 8~mm~diameter parallel-plate geometry. After loading the hydrogel, the upper plate was lowered to the prescribed gap (about 1~mm). A strain sweep was first performed to identify the linear viscoelastic region, followed by a frequency sweep at a fixed strain within this region to obtain storage modulus $G'$ and loss modulus $G''$. For PEG 8000/water mixtures, a 40~mm~diameter cone-plate geometry was used. Each liquid sample was carefully loaded to avoid air bubbles and ensure complete filling of the measurement region. Frequency-sweep tests were then performed by varying the angular frequency $\omega$. To provide a low-frequency reference for comparison with the IMR and BDT results, an apparent viscosity was estimated from the oscillatory rheometer data as $\mu = G''/\omega$, where $G''$ is the measured loss modulus and $\omega$ is the angular frequency. In this work, $\mu_{\mathrm{rheo}}$ was calculated using the measured $G''$ at $\omega=1~\mathrm{rad/s}$.
The oscillatory shear plate rheometer results are summarized in Figs.~\ref{fig: PAAm_Alginate_rheometer_results}-\ref{fig: PEG8K_rheometer_results}.

%%%%%%%%%%%%%%%%%%%%%%%%%%%%%%%%%%%%%%%%%%%%
%%%%%%%%%%%%%%%%%%%%%%%%%%%%%%%%%%%%%%%%%%%%
%%%%%%%%%%%%%%%%%%%%%%%%%%%%%%%%%%%%%%%%%%%%
%%%%%%%%%%%%%%%%%%%%%%%%%%%%%%%%%%%%%%%%%%%%

\section{Bubble Contents Dynamics}\label{app: bubble contents}
This appendix summarizes the governing equations for the physics of the bubble contents based on established literature concerning laser-induced cavitation bubbles \citep{nigmatulin1981dynamics,Akhatov2001hy,barajas2017effects,estrada2018high}, with a more comprehensive discussion available in Estrada et al.\citep{estrada2018high}. We model the bubble internal contents as a homobaric (uniform pressure), two-phase mixture of water vapor and a non-condensable gas, both treated as ideal gases.
The evolution of the (dimensionless) vapor mass fraction field, $k(r,t)$, and the temperature field, $T(r,t)$, inside the bubble $(0\leq r \leq R)$ are governed by:
\begin{equation}\label{eq:k}
\dfrac{\partial k}{\partial t} + v_{\rm m}\dfrac{\partial k}{\partial r} = \dfrac{1}{\rho_{\rm m}r^2}\dfrac{\partial }{\partial r}\left(\rho_{\rm m}r^2 D\dfrac{\partial k}{\partial r}\right)\end{equation}
and
\begin{multline}\label{eq:temp}
\dfrac{\kappa}{\kappa-1}\dfrac{p_{\rm b}}{T}\left(\dfrac{\partial T}{\partial t} + v_{\rm m}\dfrac{\partial T}{\partial r}\right) = \dot{p}_{\rm b} + \dfrac{1}{r^2}\dfrac{\partial }{\partial r}\left(r^2 K\dfrac{\partial T}{\partial r}\right) \\ + \dfrac{\kappa}{\kappa-1}\dfrac{p_{\rm b}}{T}\dfrac{{\cal R}_{\rm v}-{\cal R}_{\rm g}}{{\cal R}} D\dfrac{\partial k}{\partial r}\dfrac{\partial T}{\partial r}.\end{multline}
In these equations, $v_m(r,t)$ is the mixture velocity:
\begin{equation}
v_{\rm m}(r,t) = \dfrac{1}{\kappa p_{\rm b}}\left[(\kappa-1)K\dfrac{\partial T}{\partial r} - \dfrac{1}{3} r\dot{p}_{\rm b}\right] + \dfrac{{\cal R}_{\rm v} - {\cal R}_{\rm g}}{{\cal R}}D\dfrac{\partial k}{\partial r};
\end{equation}
$\rho_{\rm m} = {p_{\rm b}}/{{\cal R}T}$ is the mixture density field; ${\cal R} = k{\cal R}_{\rm v} + (1-k){\cal R}_{\rm g}$ is the mixture gas constant field; and $K(T) = AT+B$ is the temperature-dependent thermal conductivity of the mixture\citep{prosperetti1988nonlinear}, where $A$ and $B$ are empirical constants. The gas constants for water vapor and non-condensable gas are ${\cal R}_{\rm v}$ and ${\cal R}_{\rm g}$, respectively. The evolution of the bubble pressure $p_{\rm b}(t)$ is described by the ordinary differential equation:
\begin{multline}
\dot{p}_{\rm b} = \dfrac{3}{R}\left[-\kappa p_{\rm b} \dot{R} + (\kappa - 1)K(T|_{r=R})\left.\dfrac{\partial T}{\partial r}\right|_{r=R} \right.\\ \left. + \kappa p_{\rm b}\dfrac{{\cal R}_{\rm v}}{{\cal R}(k|_{r=R})}\dfrac{D}{1-k|_{r=R}}\left.\dfrac{\partial k}{\partial r}\right|_{r=R}\right].
\end{multline}
The constant parameters governing the bubble contents—including the binary diffusion coefficient ($D$), the specific heat ratio ($\kappa$), gas constants (${\cal R}_{\rm v}$, ${\cal R}_{\rm g}$ ), and empirical thermal conductivity constants ($A$, $B$)—are summarized in Table~\ref{tbl: material parameters}.\\

\begin{table}[!t]
\small
  \caption{\ Parameters describing the physics of the bubble contents}\label{tbl: material parameters}
  \begin{tabular*}{0.48\textwidth}{@{\extracolsep{\fill}}llll}
    \hline
     \bf{Parameter} & \bf{Value} & \bf{Parameter} & \bf{Value} \\
    \hline
%    $\rho$ & $1060$ kg/m$^3$ & $c_{long}$ & $1430$ m/s \\
%    $P_{\infty}$ & $101.3$ kPa & $\gamma$ & $5.6\times 10^{-2}$ N/m \\
    $D$ & $24.2\times10^{-6}$\,m$^2$/s & $\kappa$ & $1.4$ \\
    ${\cal R}_{\rm v}$ & $0.462$\,kJ/kg$\cdot$K & ${\cal R}_{\rm g}$ & $0.287$\,kJ/kg$\cdot$K \\
    $A$ & $5.3\times10^{-5}$\,W/m$\cdot$K$^2$ & $B$ & $1.17\times10^{-2}$\,W/m$\cdot$K \\
    $p_{\rm ref}$ & $1.17\times 10^8$\,kPa & $T_{\rm ref}$ & $5200$\,K \\
    $T_{\infty}$ & $298.15$\,K\\
    \hline
  \end{tabular*}
\end{table}

\noindent \textbf{Boundary and Initial Conditions:}\\

At the bubble center ($r = 0$), we apply symmetry boundary conditions: 
$\partial T / \partial r |_{r = 0} = \partial k / \partial r |_{r = 0} = 0$.  
At the bubble wall ($r = R$), we assume the surrounding medium is cold and isothermal, imposing the boundary condition $T|_{r=R} = T_\infty$. The partial pressure of the water vapor at the wall is assumed to equal its saturation pressure, yielding the following boundary condition for the vapor mass fraction:
\begin{equation}\label{eq:kBC}
k|_{r=R} = \left[ 1+ \dfrac{{\cal R}_{\rm v}}{{\cal R}_{\rm g}}\left(\dfrac{p_{\rm b}}{p_{\rm v,sat}(T_\infty)} - 1\right) \right]^{-1},
\end{equation}
where $p_{\rm v,sat}(T) = p_{\rm ref}\exp(-T_{\rm ref}/T)$ is the temperature-dependent saturation pressure of the water vapor with empirical constants $p_{\rm ref}$ and $T_{\rm ref}$\citep{barajas2017effects}. 
Quantitative values for $p_{\rm ref}$, $T_{\rm ref}$, and $T_\infty$ used in this study are included in Table~\ref{tbl: material parameters}.

For the initial conditions ($t = 0$), we assume the bubble contents are in thermal equilibrium. The initial temperature is therefore uniform at the ambient temperature, $T(r,t=0)=T_\infty$. Consequently, the initial vapor mass fraction, $k(r,t=0)$, is also spatially uniform and given by Eq.~\eqref{eq:kBC}. The initial bubble pressure for LIC is given by:
\begin{equation}
p_{\rm b}(t=0) = p_{\rm v,sat}(T_\infty) + \left(p_\infty + \dfrac{2\gamma}{R_0} - p_{\rm v,sat}(T_\infty)\right)\left(\dfrac{R_0}{R_{\rm max}}\right)^3.
\end{equation}
%
% For UIC, the initial bubble pressure is $p_{\rm b}(t=0) = p_\infty + {2\gamma}/{R_0}$.

\noindent \textbf{Numerical Implementation:} \\

For the numerical solution, the governing equations are first non-dimensionalized as detailed in Refs.~\cite{estrada2018high,yang2020extracting,tzoumaka2023modeling}. The spatial domain within the bubble is discretized onto a grid of $N+1=501$ equidistant points. Spatial derivatives in Eqs.~\eqref{eq:k} and \eqref{eq:temp} are then approximated using second-order central differences \citep{prosperetti1988nonlinear}. The resulting system of ordinary differential equations is evolved forward in time, in conjunction with the Keller-Miksis equation, using a variable-step, variable-order solver (\textsc{MATLAB} {\tt ode15s}).

% =================================
\section*{References}
% =================================
\bibliographystyle{unsrt}
\bibliography{reference}

\end{document}